\RequirePackage{fix-cm}
\documentclass[12pt]{article}                     
\usepackage{graphicx}
\usepackage{hyperref}
\usepackage{amsmath}
\usepackage{amsthm}
\usepackage{amsfonts}
\usepackage{hyperref}
\usepackage[top=1in, bottom=1in, left=0.5in, right=0.5in]{geometry}
\usepackage{cite}
\numberwithin{equation}{section}
\begin{document}

\title{A local phase space stochastic quantization?
}


\author{Can G\"{o}kler\footnote{gokler@fas.harvard.edu}}



\date{\vspace{-4ex}}

\maketitle

\abstract
I examine whether Nelson's stochastic formulation of Schr\"{o}dinger equation could be derived from a phase space process through a colored noise smoothing. If this conjecture is true, it would yield a local stochastic hidden variable theory. I discuss how this does not necessarily contradict Bell type theorems as general local stochastic theories can violate local causality assumptions. I also discuss the generalization to quantization of fields and speculate about the gravitational origins of noise.


\tableofcontents

\newpage

\section{Introduction}

Ever since the advent of quantum theory, the question of whether there exists a local deterministic or stochastic theory underlying quantum mechanics has been of considerable interest. One important direction was provided by Nelson's seminal work on the reformulation of quantum mechanics in terms of stochastic mechanics of particles \cite{Nelson66, Nelson1, Nelson2}. Stochastic mechanics is an exact reformulation of quantum theory like the path integral formulation. However, when applied to multiple particles and fields, it is manifestly non-local. Nelson suggested the important question \cite{Nelson2}: Can one give a local model which would reduce to stochastic mechanics in an appropriate limit?. This is the central question, I am concerned with in this note. Although, it is folklore to believe that it is impossible, I argue following Nelson that Bell type theorems do not actually rule out local stochastic theories which in general can violate local causality assumptions. I outline the construction of a simple local theory and formulate conjectures which would enable such a theory to exist first for the single particle case, then for multiple particles and finally for the prototypical example of the scalar field. Of course, existence of such a theory does not mean that the universe obeys it. Assuming that the theory is physical, I will also speculate about potential gravitational origins of noise. In the absence of evidence for the quantization of the gravitational field, this could provide a consistent view of the universe where quantum theory arises out of gravitational interactions for all the fields in the universe. This is similar to that Brownian motion of a single particle arises from the microscopic Newtonian dynamics of a gas.

To see the main difficulty, consider the Madelung formulation of quantum theory:

\begin{equation}
\frac{\partial \rho}{\partial t} = - \frac{\partial}{\partial x} ( \rho \frac{1}{m} \frac{\partial S}{\partial x} )
\end{equation}

\begin{equation}
\frac{\partial S}{\partial t} = - \frac{1}{2m} ( \frac{\partial S}{\partial x} )^2 - U(x) + \frac{\hbar^2}{2m} \frac{1}{\sqrt{\rho}} \frac{\partial^2}{\partial x^2} \sqrt{\rho}
\end{equation}
where $\rho(x,t)$ is the probability of finding the particle at $(x,t)$ and $S(x,t)$ is the phase of the wave function. The second equation would be the classical Hamilton-Jacobi equation if the last term (quantum potential) was absent, and the equations would describe an ensemble of classical particles. Thus, a derivation of Schr\"{o}dinger equation must somehow generate the quantum potential term either implicitly or explicitly. Nelson's stochastic mechanics in terms of particles trajectories (see section \ref{Nelson1section}) provides an implicit understanding of this term in terms of averaged stochastic derivatives. If there are multiple particles, the quantum potential term in general depends on all the positions of the particles. A stochastic unraveling such as Nelson's theory then becomes non-local. The central question is equivalent to asking whether one can induce the quantum potential term while retaining locality. Nelson argued that Markovianity of his theory might be the main obstacle against formulating a local underlying theory and a non-Markovian theory is not necessarily non-local in \cite{Nelson2}. I will propose a way to circumvent this problem through replacing the white noise assumption in Nelson's theory with colored noise. One important objection against the existence of a such a local theory comes from believes surrounding Bell-type theorems or inequalities. These theorems assumes two kinds of locality: the usual locality in the sense of relativity and local causality. Local causality assumptions do not have to be satisfied by general local stochastic theories. I will discuss these issues in section \ref{localitysection}.

The outline is as follows. In section \ref{sectionsingleparticle}, I give an overview of Nelson's theory for a single particle then propose a phase space formulation based on colored noise. I formulate conjectures related to the existence of such a theory. In section \ref{section multiple partciles}, I generalize the arguments given for the single particle case to the multiple particle case. In section \ref{localitysection}, I elaborate on the issues of locality and local causality. In section \ref{sectionfields}, I generalize the ideas to the case of the scalar field and speculate about the gravitational origin of noise. 

\section{Single particle} \label{sectionsingleparticle}
\subsection{Nelson's stochastic mechanics for a single particle} \label{Nelson1section}

\indent

I give a brief review of Nelson's stochastic formulation of non-relativistic quantum mechanics in one dimension. For more details see Nelson's original paper\cite{Nelson66}, his two books\cite{Nelson1, Nelson2} and Guerra's review\cite{Guerra}. Consider the Schr\"{o}dinger equation:

\begin{equation}
i \hbar \frac{\partial \psi(x,t)}{\partial t} = (-\frac{\hbar^2}{2m} \frac{\partial^2}{\partial x^2} + U(x)) \psi(x,t).
\end{equation}
Putting $\psi(x,t) = \sqrt{\rho(x,t)} e^{\frac{i}{\hbar} S(x,t)}$ we get the Madelung equations:

\begin{equation}
\frac{\partial \rho}{\partial t} = - \frac{\partial}{\partial x} ( \rho \frac{1}{m} \frac{\partial S}{\partial x} )
\end{equation}

\begin{equation}
\label{Madelung2}
\frac{\partial S}{\partial t} = - \frac{1}{2m} ( \frac{\partial S}{\partial x} )^2 - U(x) + \frac{\hbar^2}{2m} \frac{1}{\sqrt{\rho}} \frac{\partial^2}{\partial x^2} \sqrt{\rho}
\end{equation}
where $\rho(x,t)$ is the probability of finding the particle at $(x,t)$ and $S(x,t)$ is the phase of the wave function. We recognize first of the equations as the continuity equation with velocity $\frac{1}{m} \frac{\partial S}{\partial x}$. The second of the equations apart from the last term (quantum potential) on the right hand side is the Hamilton-Jacobi equation. Thus if $\hbar = 0$, we have the classical ensemble of particles. The Newton's equations of motion are then the equations that characteristic curves obey corresponding to this set of Madelung partial differential equations. Since the quantum potential term depends on the probability $\rho(x,t)$, giving deterministic characteristics seems not possible. However as Nelson proved\cite{Nelson66, Nelson1, Nelson2}, it is possible to give a Markovian stochastic process associated to the solution of Madelung equations in position space. We start by assuming that a particle obeys the following stochastic differential equation:

\begin{equation} \label{Nelsonfirst}
 dx(t) = b(x(t), t) dt + \sqrt{\frac{\hbar}{m}} dW(t)
\end{equation}
where $b(x(t),t)$ is a general function and $dW$ is the Wiener process. We will always interpret the stochastic differential equations in this paper in It\^{o} sense. We call this as Nelson's first postulate. The diffusion equation associated to this is\cite{Friedman, Arnolds}

\begin{equation}
\frac{\partial \rho(x,t)}{\partial t} = - \frac{\partial}{\partial x} ( b(x,t) \rho(x,t) ) +  \frac{\hbar}{2m} \frac{\partial^2}{\partial x^2}\rho(x,t)
\end{equation}
where $\rho(x,t)$ is the probability of finding the particle at $x$ at time $t$. In order to match with the continuity equation we define 

\begin{equation}
\frac{\partial}{\partial x}  S(x,t)= m ( b(x,t) - \frac{\hbar}{2m} \frac{\partial}{\partial x } \log \rho(x,t) )
\end{equation}
where we assumed that $\rho(x,t)$ is nowhere zero. For a discussion of what happens at zeros see \cite{Nelson2} and references therein. Indeed, the probability of the particle arriving at places where $\rho(x,t)=0$ is zero since $b(x,t)$ is singular at zeros and acts repulsively on the stochastic particle trajectory pushing the particle away from the zeros. We want $S(x,t)$ just defined in this way to satisfy the quantum Hamilton-Jacobi equation. We could postulate it as a partial differential equation but Nelson found a way to write this solely in terms of the stochastic particle trajectory. The quantum Hamilton-Jacobi equation can be shown to be equivalent to the following equation:

\begin{equation}
\frac{1}{2}(D_+D_- + D_- D_+ ) x(t) = -\frac{1}{m} \frac{\partial U(x)}{\partial x} |_{x(t)}
\end{equation}
where $D_+$ and $D_-$ are forward and backward derivatives which will be defined below, the right hand side is the classical acceleration of the particle  evaluated on the stochastic trajectory and the left hand side is the time-symmetric stochastic acceleration. This is the stochastic analogue of Newton's second law. Thus we call this as Newton-Nelson law or Nelson's second law. The forward and backward derivatives are defined to be

\begin{equation}
D_+ x(t) = \lim_{\Delta t \to 0^+}  E [\frac{x(t+ \Delta t) - x(t)}{\Delta t} | x(t)]
\end{equation}

\begin{equation}
D_- x(t) =\lim_{\Delta t \to 0^+}  E [\frac{x(t) - x(t - \Delta t)}{\Delta t} | x(t)]
\end{equation}
where $E[f| x(t) ]$ denotes the expectation of $f$ conditioned on $x(t)$. For any function $F(x,t)$ we can write its forward and backward derivatives explicitly as follows

\begin{equation} \label{D+}
(D_+F)(x,t) = \frac{\partial}{\partial t} F(x,t) + b(x,t) \frac{\partial}{\partial x} F(x,t) + \frac{\hbar}{2m} \frac{\partial^2}{\partial x^2} F(x,t)
\end{equation}

\begin{equation} \label{D-}
(D_-F)(x,t) = \frac{\partial}{\partial t} F(x,t) + (b(x,t)-\frac{\hbar}{m} \frac{\partial}{\partial x} \log \rho(x,t)) \frac{\partial}{\partial x} F(x,t) - \frac{\hbar}{2m} \frac{\partial^2}{\partial x^2} F(x,t).
\end{equation}
The derivation of the formula for $D_+$ is straightforward but the calculation of $D_-$ is subtler\cite{Nelson1, Nelson2, Guerra}. Using these formulas it is straightforward to show that the Newton-Nelson law is equivalent to the $x$ derivative of the second Madelung equation (eq.$\ref{Madelung2}$). It has been shown that for each solution of the Schr\"{o}dinger equation there is an associated stochastic process satisfying Nelson's postulates and if Nelson's postulates are satisfied that one can construct a wave function which satisfies the Schr\"{o}dinger equation with its absolute square the probability density of the position of particle. The stochastic formulation can be generalized to particles propagating in higher dimensions, multiple particles, fields and particles with spin\cite{Nelson2, Guerra}.

\subsection{Colored noise smoothed quantum process}

Let $A_\beta$ be the colored noise defined by the equation

\begin{equation} \label{color}
dA_\beta = -\beta A_\beta dt + \beta dW
\end{equation}
where $dW$ is the Wiener process. Heuristically as $\beta \rightarrow \infty$, the left hand side can be neglected yielding

\begin{equation}
A_\beta \rightarrow \frac{dW}{dt}.
\end{equation}
After the transient time $t \gg 1/\beta$, 
$\langle A_\beta(t) \rangle \approx 0 $ and $\langle A_\beta(t) A_\beta(t')\rangle \approx \beta \exp(\beta |t-t'|)$ where $\langle \cdot \rangle $ denotes expactation. Thus, $\tau=1/\beta$ is the correlation time. If $x_\beta$ is driven by $A_\beta$ as in

\begin{equation}
dx_\beta = b(x_\beta ,t)dt + \epsilon A_\beta dt
\end{equation}
then as $\beta \rightarrow \infty$, $x_\beta$ converges to the solution of 

\begin{equation}
dx = b(x, t) dt + \epsilon dW 
\end{equation}
where $x_\beta(0)=x(0)$. This can be made mathematically rigorous \cite{Nelson1, Pavliotis} Although more general colored noises are possible, I will restrict to those described by eq. \ref{color} in this note for simplicity.

Suppose a quantum process $x_q(t)$ is associated with a solution of Schr\"{o}dinger equation through Nelson's formulation. An equivalent characterization is the pair $(b(x,t), \rho(x,0))$ (one can solve for $x_q(t)$ using eq. \ref{Nelsonfirst}). Now replace the white noise in eq. \ref{Nelsonfirst} with colored noise $A_\beta$ to define the new process 

\begin{equation}
dx(t) = b(x(t), t ) dt + \epsilon A_\beta(t)dt
\end{equation}
where $\epsilon = \sqrt{\hbar / m}$, with the same initial density $\rho(x, 0)$. As $\beta \rightarrow \infty$, this should be a good approximation to $x_q(t)$. However the stochastic acceleration of $x(t)$ will violate Nelson's second law:

\begin{equation}
\frac{1}{2}(D_+D_- + D_- D_+ ) x(t) \neq -\frac{1}{m} \frac{\partial U(x)}{\partial x} |_{x(t)}
\end{equation}
as can be seen by direct calculation (see next section). The violation of Nelson's second law is of the same order of the quantum potential. As $x(t)$ and $x_q(t)$ are close to each other for large $\beta$ in an appropriate norm, their stochastic derivatives differ. Of course, this is common as even two deterministic functions with wildly differing derivatives can be close in norm (e.g. take a reasonably smooth function and consider a function rapidly fluctuating around it). This is why I call $x(t)$ as a colored noise smoothing of $x_q(t)$.

If on the other hand, we consider the white noise limit of $A_\beta$ and calculate small correlation time $\tau=1/\beta$ expansion, we get a process of the form \cite{Jung}

\begin{equation}
dx_\beta(t) = b(x_\beta(t), t) dt +  \epsilon (1 + \tau g(x_\beta(t), t)) dW(t)
\end{equation}
for some $g(x, t)$ related to $b(x,t)$. Now we see that for large $\beta$, $x_\beta$ approximately satisfies Nelson's second law and it is also close to $x_q(t)$. The singular nature of the white noise limit is evident as $D_+$ and $D_-$ are sensitive to the presence of Wiener terms as can be seen from the definitions (eqs. \ref{D+} and \ref{D-}).

\subsection{Calculation of stochastic acceleration for a colored noise process}

Consider the process of the form for given $b(x,t)$:

\begin{align}
&dx(t) = b(x(t), t )dt + \epsilon A_\beta (t) dt \\ \nonumber
&dA_\beta (t) = -\beta A_\beta (t) dt + \beta dW(t)
\end{align}
where initially $x$ and $A_\beta$ are uncorrelated. We will make use of the following formulas for forward and backward derivatives conditioned on fixed $(x(t), A_\beta(t) )$ of a function $G(x,A_\beta,t)$ which can be found in section 5 of Guerra's review\cite{Guerra}:

\begin{align} \label{D+phase}
(D_+ G)(x,A_\beta,t) |_{(x(t), A_\beta(t))} &= \lim_{\Delta t \to 0^+} E[ \frac{G(x(t+\Delta t), A_\beta(t + \Delta t), t+\Delta t) - G(x(t), A_\beta(t), t)}{\Delta t} | x(t), A_\beta(t)]
\\ \nonumber &= \frac{\partial G}{\partial t} + (b +\epsilon A_\beta ) \frac{\partial G}{\partial x} - \beta A_\beta \frac{\partial G}{\partial A_\beta} + \frac{\beta^2}{2} \frac{\partial^2G}{\partial A_\beta^2}
\end{align}

\begin{align} \label{D-phase}
 (D_- G)(x,A_\beta,t) |_{(x(t), A_\beta(t))}  &= \lim_{\Delta t \to 0^+} E[ \frac{G(x(t), A_\beta(t), t) - G(x(t-\Delta t), A_\beta(t - \Delta t), t-\Delta t) }{\Delta t} | x(t), A_\beta(t)] \\ \nonumber &= \frac{\partial G}{\partial t} + (b+\epsilon A_\beta) \frac{\partial G}{\partial x} + ( -\beta A_\beta - \beta^2 \frac{\partial}{\partial A_\beta} \log \rho(x,A_\beta,t) )  \frac{\partial G}{\partial A_\beta} - \frac{\beta^2}{2} \frac{\partial^2G}{\partial A_\beta^2}
\end{align}
We also need the following result on conditional expectations for a set of random variables $(x,A,z)$:

\begin{equation} \label{expectation1}
E [ F(z) | x] = \int E [F(z) | A, x] p (A| x) dA
\end{equation}
for any function $F(z)$. 
We see that $D_+ x = D_- x |_{(x, A_\beta)}= b+A_\beta$. Then

\begin{equation}
D_-D_+ x |_{(x, A_\beta)} = D_- (b+A_\beta) |_{(x, A_\beta)} = \frac{\partial b}{\partial t}  + (b +\epsilon A\beta) \frac{\partial b}{\partial x}  -\beta A_\beta  - \beta^2 \frac{\partial}{\partial A_\beta} \log \rho(x,A_\beta,t)  
\end{equation}
and 
\begin{equation}
D_+D_- x |_{(x, A_\beta)} = D_+(b+A_\beta) |_{(x, A_\beta)} =\frac{\partial b}{\partial t}  + (b +\epsilon A_\beta) \frac{\partial b}{\partial x}  -\beta A_\beta 
\end{equation}
Using eq. \ref{expectation1}, we calculate the stochastic acceleration of $x(t)$ as

\begin{equation}
\frac{1}{2} (D_+ D_- + D_- D_+ ) x(t) |_{x(t)}= \frac{\partial b}{\partial t} + b \frac{\partial b}{\partial x} + (\epsilon \frac{\partial b}{\partial x} - \beta) \int A_\beta \rho_t(A_\beta | x) dA_\beta
\end{equation}
For large $\beta$, $A_\beta$ is close to white noise so we can assume $\rho_t(A_\beta | x) \approx \rho_t(A_\beta)$ and neglect the last term after the transient time as $\langle A_\beta \rangle \approx 0$. Thus we see that Nelson's second law is violated if $b(x,t)$ corresponds to a quantum process.

\subsection{A phase space derivation?}

Now consider the phase space process 

\begin{align} \label{phaseprocess}
&dx_\beta(t) = v_\beta(t) dt + \epsilon A_\beta(t) dt \\ \nonumber	
&dv_\beta(t) = a(x_\beta(t)) dt \\ \nonumber
&dA_\beta(t) = -\beta A_\beta(t) dt + \beta dW(t)
\end{align}
where $v_\beta(t)$ is the velocity of the particle and $a(x_\beta(t))$ is the acceleration on the particle. One can show through a similar calculation as last section that Nelson's second law is satisfied by $x_\beta(t)$ for large but finite $\beta$. Also for large $\beta$, integrating out $v_\beta$ and $A_\beta$, $x_\beta(t)$ approximately satisfies

\begin{equation} \label{reducedequation}
dx_\beta(t) = (\int v \rho_t(v|x_\beta)dv ) dt + \epsilon A_\beta(t) dt
\end{equation}
where we identify $ b(x,t) = \int v \rho_t(v|x)dv$ as the drift velocity. Hence, $x_\beta(t)$ also approximately satisfies Nelson's first law. Can we say that as $\beta$ goes large then $x_\beta(t)$ becomes a quantum process satisfying both Nelson's first and second law exactly? More precisely, given $x_\beta(t)$ satisfying eq. \ref{phaseprocess} for large $\beta$, does there exist a quantum process $x_q(t)$ near $x_\beta(t)$? Conversely for a given quantum process, does there exist $x(t)$ near it satisfying eq. \ref{phaseprocess}? I don't know the answers to these questions. Here is the subtlety similar to that we encountered in the colored noise smoothing of a quantum process. In the singular limit when we replace $A_\beta(t)$ by white noise we have

\begin{align}
&dx_\infty(t) = v_\infty(t) dt + \epsilon dW(t) \\ \nonumber
&dv_\infty(t) = a(x_\infty(t))dt
\end{align}
If we calculate the stochastic acceleration we see that due to the presence of the Wiener term, extra terms are induced. Thus $x_\infty(t)$ violates the second law. We can trace this back to the ordering of limits $\Delta t \rightarrow 0$ in the definition of $D_+$ and $D_-$ and $\beta \rightarrow \infty$. In general

\begin{equation}
 (D_+ D_- + D_- D_+) \lim_{\beta \rightarrow \infty} x_\beta(t) \neq \lim_{\beta \rightarrow \infty} (D_+ D_- + D_- D_+) x_\beta(t).
\end{equation}

What have we learned until now? We know that given a quantum process we can construct a colored noise smoothing close to it albeit violating Nelson's second law. If we make a correlation time expansion around the singular white noise limit, we get a process close to the quantum process and approximately satisfying Nelson's second law. The phase space process on the other hand exactly satisfies Nelson's second law for all finite and large $\beta$ and approximates Nelson's first law. However in the singular limit, it fails to satisfy Nelson's second law. Therefore it is not obvious whether the questions raised above have affirmative answers. Hence I formulate two conjectures:

\indent

\textbf{Conjecture 1.} Let $(x(t), v(t), A_\beta(t))$ satisfy eq. \ref{phaseprocess}. Assume that $A_\beta$ is initially uncorrelated to $x$ and $v$. Then as $\beta$ becomes large, there exists a quantum process $x_q(t)$, satisfying both of Nelson's laws (equivalent to a solution of the Schr\"{o}dinger equation), which is close to $x(t)$ projection of the phase space process in some appropriate norm. 

\indent

\textbf{Conjecture 2.} Converse of Conjecture 1. Let $x_q(t)$ be a quantum process. Then there exist a phase space process of the form given by eq. \ref{phaseprocess}, which is close to $x_q(t)$ in some appropriate norm, as $\beta$ becomes large.

I will argue in the following section that Conjecture 1 implies Conjecture 2.

\subsection{Conjecture 1 implies Conjecture 2}

Given a solution $\psi(x,t)$ to the Schr\"{o}dinger equation, split it into probability and phase functions as $\psi(x,t)=\sqrt{\rho(x,t)}\exp(\frac{i}{\hbar} S(x,t))$. Then $\rho(x,t)$ and $S(x,t)$ satisfy the Madelung equations:

\begin{equation}
\frac{\partial \rho}{\partial t} = - \frac{\partial}{\partial x} ( \rho \frac{1}{m} \frac{\partial S}{\partial x} )
\end{equation}

\begin{equation}
\frac{\partial S}{\partial t} = - \frac{1}{2m} ( \frac{\partial S}{\partial x} )^2 - U(x) + \frac{\hbar^2}{2m} \frac{1}{\sqrt{\rho}} \frac{\partial^2}{\partial x^2} \sqrt{\rho}
\end{equation}
We define $b(x,t)$ as

\begin{equation} \label{bdef}
 b(x,t) = \frac{1}{m}\frac{\partial}{\partial x}  S(x,t)+\frac{\hbar}{2m} \frac{\partial}{\partial x } \log \rho(x,t).
\end{equation}
Then $\rho(x,t)$ satisfies 

\begin{equation} \label{NelsonFP}
\frac{\partial \rho(x,t)}{\partial t} = - \frac{\partial}{\partial x} ( b(x,t) \rho(x,t) ) +  \frac{\hbar}{2m} \frac{\partial^2}{\partial x^2}\rho(x,t)
\end{equation}
which is equivalent to Nelson's first postulate:

\begin{equation} 
 dx(t) = b(x(t), t) dt + \sqrt{\frac{\hbar}{m}} dW(t).
\end{equation}
Therefore given a solution to the Schr\"{o}dinger equation, the associated Nelson process is completely specified by $\rho(x,t)$ and $b(x,t)$.  We would like to construct an approximate solution to eq.\ref{phaseprocess} to match with the Fokker-Planck equation (eq.$\ref{NelsonFP}$). Since $\rho(x,t)$ is already fixed by the solution to the Schr\"{o}dinger equation, the only free function is $\hat{\rho}(v|x,t)$ as these two determine the joint density $\hat{\rho}(x,v,t)=\hat{\rho}(v | x, t)\rho(x,t)$. The mean of $\hat{\rho}(v|x,t)$ is fixed by $b(x,t)$ (eq.\ref{reducedequation}) which is determined from the solution of the Schr\"{o}dinger equation via eq.$\ref{bdef}$. Furthermore, we have the natural probability normalization constraint $\int \hat{\rho}(v | x,t) dv = 1$. Therefore, the problem becomes the following: Given a solution to the Schr\"{o}dinger equation represented by $\rho(x,t)$ and $b(x,t)$, construct an approximate solution $\hat{\rho}(x,v,t)$ (construct $\hat{\rho}(v| x, t)$)  to eq.$\ref{phaseprocess}$ such that $\hat{\rho}(x,v,t)=\hat{\rho}(v|x,t)\rho(x,t)$, $b(x,t) = \int v \hat{\rho}(v|x, t) dv$ and $\int \hat{\rho}(v | x,t) dv = 1$. To construct a solution, choose the initial conditional density $\hat{\rho}(v|x, t=0)$ such that 

\begin{equation} \label{velocityav}
\int v \hat{\rho}(v| x, t=0) dv = b(x, t=0).
\end{equation}
Assuming conjecture 1 is true, the solutions to eq.$\ref{phaseprocess}$ is close to $x_q(t)$ satisfying Nelson's both laws. Therefore, eq.$\ref{phaseprocess}$ approximately propagates the initial $(\rho(x, t=0), b(x, t=0))$ to $(\hat{\rho}(x,t), \hat{b}(x,t))$ according to the Schr\"{o}dinger equation. Since $(\rho(x, t=0), b(x, t=0))$ completely specifies the initial conditions for the Schr\"{o}dinger equation (up to an additive constant in the initial phase $S(x,t=0)$), the solutions are unique. Therefore $(\hat{\rho}(x,t), \hat{b}(x,t))=(\rho(x,t), b(x,t))$. Hence given a solution to the Schr\"{o}dinger equation specified by $\rho(x,t)$ and $b(x,t)$, we see that any initial density $\hat{\rho}(x, v, t=0)$ satisfying eq.$\ref{velocityav}$ and $\int \hat{\rho}(x, v, t=0)dv=\rho(x,t=0)$ yields an approximate solution $\hat{\rho}(x, v, t)$ to eq.\ref{phaseprocess} such that $\int v \hat{\rho}(v| x, t) dv = b(x, t)$ and $\int \hat{\rho}(x, v, t)dv=\rho(x,t)$.  Note that there are infinitely many ways to choose the initial conditional density $\hat{\rho}(v| x, t=0)$ compatible with eq.\ref{velocityav}.

\section{Multiple particles} \label{section multiple partciles}
\subsection{Nelson's mechanics for multiple particles}

We have worked with a single particle until now. In this section, I give a review of Nelson's stochastic formulation of non-relativistic quantum mechanics for $n$ particles with position variables $\vec{x}=(x_1,...,x_n)$ and masses $m_1, ..., m_n$, interacting via the potential $U(\vec{x})$. Consider the Schr\"{o}dinger equation:

\begin{equation}
i \hbar \frac{\partial \psi(\vec{x},t)}{\partial t} = (-\sum_{j=1}^n \frac{\hbar^2}{2m_j} \frac{\partial^2}{\partial x_j^2} + U(\vec{x})) \psi(\vec{x},t).
\end{equation}
Putting $\psi(\vec{x},t) = \sqrt{\rho(\vec{x},t)} e^{\frac{i}{\hbar} S(\vec{x},t)}$, we get the Madelung equations:

\begin{equation}
\frac{\partial \rho(\vec{x},t)}{\partial t} = - \sum_{j=1}^n \frac{\partial}{\partial x_j} ( \rho(\vec{x},t) \frac{1}{m_j} \frac{\partial S(\vec{x},t)}{\partial x_j} )
\end{equation}

\begin{equation}
\label{Madelung21}
\frac{\partial S(\vec{x},t)}{\partial t} = - \sum_{j=1}^n\frac{1}{2m_j} ( \frac{\partial S(\vec{x},t)}{\partial x_j} )^2 - U(\vec{x}) + \sum_{j=1}^n\frac{\hbar^2}{2m_j} \frac{1}{\sqrt{\rho(\vec{x},t)}} \frac{\partial^2}{\partial x_j^2} \sqrt{\rho(\vec{x},t)}
\end{equation}
where $\rho(\vec{x},t)$ is the probability of finding the particles at $\vec{x}$ at time $t$ and $S(\vec{x},t)$ is the phase of the wave function. We recognize first of the equations as the continuity equation with velocity $(\frac{1}{m_1} \frac{\partial S}{\partial x_1}, ..., \frac{1}{m_n} \frac{\partial S}{\partial x_n} )$. The second of the equations apart from the last term (quantum potential) on the right hand side is the Hamilton-Jacobi equation. Thus if $\hbar = 0$, we have the classical ensemble of particles. We start by assuming that a particles obey the following stochastic differential equations:

\begin{equation} \label{Nelsonfirst12}
 dx_i(t) = b_i(\vec{x}(t), t) dt + \sqrt{\frac{\hbar}{m_i}} dW_i(t)
\end{equation}
where each $b_i(\vec{x},t)$ is a general function of positions and time, and $(dW_1, ..., dW_n)$ are independent Wiener processes. We call this as Nelson's first law or postulate. The diffusion equation associated to this is\cite{Friedman, Arnolds}

\begin{equation}
\frac{\partial \rho(\vec{x},t)}{\partial t} = - \sum_{j=1}^n\frac{\partial}{\partial x_j} ( b_j(\vec{x},t) \rho(\vec{x},t) ) +   \sum_{j=1}^n\frac{\hbar}{2m_j} \frac{\partial^2}{\partial x_j^2}\rho(\vec{x},t)
\end{equation}
where $\rho(\vec{x},t)$ is the probability of finding the particles at $\vec{x}$ at time $t$. In order to match with the continuity equation, we define 

\begin{equation} \label{gradient1}
\frac{\partial}{\partial x_i}  S(\vec{x},t)= m_i ( b_i(\vec{x},t) - \frac{\hbar}{2m_i} \frac{\partial}{\partial x_i } \log \rho(\vec{x},t) )
\end{equation}
where we assumed that $\rho(\vec{x},t)$ is nowhere zero. For a discussion of what happens at zeros, see Section 2. Note that in eq.\ref{gradient1}, we assume that $(b_1,...,b_n)$ is a gradient. At first sight, this seems like an independent assumption. However, it can be shown that the gradient assumption is implied by the Newton-Nelson law using a variational argument (see Theorem 6 of \cite{yasue} and \cite{Nelson2}). Newton-Nelson law is introduced below. We want $S(\vec{x},t)$ just defined in this way to satisfy the quantum Hamilton-Jacobi equation. The quantum Hamilton-Jacobi equation can be shown to be equivalent to the following equations:

\begin{equation}
\frac{1}{2}(D_+D_- + D_- D_+ ) x_i(t) = -\frac{1}{m_i} \frac{\partial U(\vec{x})}{\partial x_i} |_{\vec{x}(t)}
\end{equation}
where $D_+$ and $D_-$ are forward and backward derivatives which will be defined below, the right hand side is the classical acceleration of the particle evaluated on the stochastic trajectory and the left hand side is the time-symmetric stochastic acceleration. This is the stochastic analogue of Newton's second law. Thus we call this as Newton-Nelson law or Nelson's second law. The forward and backward derivatives are defined to be

\begin{equation}
D_+ x_i(t) = \lim_{\Delta t \to 0^+}  E [\frac{x_i(t+ \Delta t) - x_i(t)}{\Delta t} | \vec{x}(t)],
\end{equation}

\begin{equation}
D_- x_i(t) =\lim_{\Delta t \to 0^+}  E [\frac{x_i(t) - x_i(t - \Delta t)}{\Delta t} | \vec{x}(t)]
\end{equation}
where $E[f| \vec{x}(t) ]$ denotes the expectation of $f$ conditioned on $\vec{x}(t)$. For any function $F(\vec{x},t)$ we can write its forward and backward derivatives explicitly as follows

\begin{equation} \label{D+1}
(D_+F)(\vec{x},t) = \frac{\partial}{\partial t} F(\vec{x},t) + \sum_{j=1}^n b_j(\vec{x},t) \frac{\partial}{\partial x_j} F(\vec{x},t) +\sum_{j=1}^n \frac{\hbar}{2m_j} \frac{\partial^2}{\partial x_j^2} F(\vec{x},t),
\end{equation}

\begin{equation} \label{D-1}
(D_-F)(\vec{x},t) = \frac{\partial}{\partial t} F(\vec{x},t) + \sum_{j=1}^n(b_j(\vec{x},t)-\frac{\hbar}{m_j} \frac{\partial}{\partial x_j} \log \rho(\vec{x},t)) \frac{\partial}{\partial x_j} F(\vec{x},t) - \sum_{j=1}^n\frac{\hbar}{2m_j} \frac{\partial^2}{\partial x_j^2} F(\vec{x},t).
\end{equation}
Using these formulas, it is straightforward to show that the Newton-Nelson law is equivalent to the $\vec{x}$ derivative of the second Madelung equation (eq.$\ref{Madelung21}$). It has been shown that for each solution of the Schr\"{o}dinger equation, there is an associated stochastic process satisfying Nelson's postulates, and if Nelson's postulates are satisfied, that one can construct a wave function which satisfies the Schr\"{o}dinger equation, with its absolute square the probability density of the positions of particles \cite{Nelson66, Nelson1, Nelson2}.

\subsection{A phase space derivation for multiple particles? }

The multi-particle case is a straightforward generalization of the single particle case. Consider the multi-particle phase space process

\begin{align} \label{phaseprocessmulti1}
&dx_i(t) = v_i (t) dt + \epsilon A_i (t) dt \\ \nonumber	
&dv_i(t) = a(x_1(t), ... ,x_n(t)) dt \\ \nonumber
&dA_i (t) = -\beta_i A_i (t) dt + \beta_i dW_i(t)
\end{align}
where $x_i$, $v_i$, $a(x_i)=-(1/m_i) \partial_{x_i} U(x_1, ...,  x_n, t)$ are the position, velocity and the acceleration of particle $i$. Like the single particle case, one can show that Newton-Nelson law is satisfied for large $\beta_i$. Also integrating out the noises and velocities, we can approximate $x_i(t)$ by 

\begin{equation} \label{reducedequation2}
dx_i(t) = (\int v_i \rho_t(v_i | x_1, ..., x_n)dv_i ) dt + \epsilon A_i dt
\end{equation}
identifying $b_i(x_1, ...., x_n, t)= \int v_i \rho_t(v_i | x_1, ..., x_n)dv_i $ as the drift velocity for particle $i$. Here we assumed that close to the white noise limit with large $\beta_i$, $A_{i_1}$ and $A_{i_2}$ become independent for $i_1 \neq i_2$. Subtleties similar to that in the single particle case remain here at the white noise limit in calculating the stochastic acceleration.  I formulate conjectures which parallel the single particle case.  

\indent

\textbf{Conjecture 3.} Let $ \{(x_i(t), v_i(t), A_i(t)) \}_i$ satisfy eq. \ref{phaseprocessmulti1}. Assume that $A_i$ is initially uncorrelated to $\vec{x}$ and $\vec{v}$. Then as $\{\beta_i\}_i$ become large, there exists a quantum process $\vec{x}_q(t)$, satisfying both of Nelson's laws (equivalent to a solution of the Schr\"{o}dinger equation), which is close to $\vec{x}(t)$ projection of the phase space process in some appropriate norm. 

\indent

\textbf{Conjecture 4.} Converse of Conjecture 1. Let $\vec{x}(t)$ be a quantum process. Then there exist a phase space process of the form given by eq. \ref{phaseprocessmulti1}, which is close to $\vec{x}_q(t)$ in some appropriate norm, as $\{\beta_i \}_i$ become large.

Next, I argue that Conjecture 3 implies Conjecture 4 with by a similar argument given for the single particle case.

\subsection{Conjecture 3 implies Conjecture 4}

Given a solution $\psi(\vec{x},t)$ to the Schr\"{o}dinger equation we split it into probability and phase functions as $\psi(\vec{x},t)=\sqrt{\rho(\vec{x},t)}\exp(\frac{i}{\hbar} S(\vec{x},t))$. Then $\rho(\vec{x},t)$ and $S(\vec{x},t)$ satisfy the Madelung equations:

\begin{equation}
\frac{\partial \rho}{\partial t} = - \sum_{i=1}^n\frac{\partial}{\partial x_i} ( \rho \frac{1}{m_i} \frac{\partial S}{\partial x_i} )
\end{equation}

\begin{equation}
\frac{\partial S}{\partial t} = -  \sum_{i=1}^n\frac{1}{2m_i} ( \frac{\partial S}{\partial x_i} )^2 - U(\vec{x}) +  \sum_{i=1}^n \frac{\hbar^2}{2m_i} \frac{1}{\sqrt{\rho}} \frac{\partial^2}{\partial x_i^2} \sqrt{\rho}
\end{equation}
where the potential is $U(x)= \sum_{i=1}^n\frac{m_i}{2}\omega_i^2 x_i^2$ for $n$ non-interacting particles. We define $b_i(x,t)$ as

\begin{equation} \label{bdef1}
 b_i(\vec{x},t) = \frac{1}{m_i}\frac{\partial}{\partial x_i}  S(\vec{x},t)+\frac{\hbar}{2m_i} \frac{\partial}{\partial x_i } \log \rho(\vec{x},t).
\end{equation}
Then $\rho(x,t)$ satisfies 

\begin{equation} \label{NelsonFP1}
\frac{\partial \rho(\vec{x},t)}{\partial t} = -  \sum_{i=1}^n\frac{\partial}{\partial x_i} ( b_i(\vec{x},t) \rho(\vec{x},t) ) +   \sum_{i=1}^n\frac{\hbar}{2m_i} \frac{\partial^2}{\partial x_i^2}\rho(\vec{x},t)
\end{equation}
which is equivalent to Nelson's first postulate:

\begin{equation} 
 dx_i(t) = b_i(\vec{x}(t), t) dt + \sqrt{\frac{\hbar}{m_i}} dW_i(t)
\end{equation}
for each $i$. Therefore, given a solution to the Schr\"{o}dinger equation, the associated Nelson process is completely specified by $\rho(\vec{x},t)$ and $(b_1(\vec{x},t),...,b_n(\vec{x},t))$.  We would like to construct an approximate solution to eq.$\ref{phaseprocessmulti}$ to match with the Fokker-Planck equation (eq.$\ref{NelsonFP1}$). Since $\rho(\vec{x},t)$ is already fixed by the solution to the Schr\"{o}dinger equation, the only free function is $\hat{\rho}(\vec{v}|\vec{x},t)$, as these two determine the joint density $\hat{\rho}(\vec{x},\vec{v},t)=\hat{\rho}(\vec{v} | \vec{x}, t)\rho(\vec{x},t)$. The mean of $\hat{\rho}(\vec{v}|\vec{x},t)$ is fixed by $(b_1(\vec{x},t),...,b_n(\vec{x},t))$, which is determined from the solution to the Schr\"{o}dinger equation via eq.$\ref{bdef1}$. Furthermore, we have the natural probability normalization constraint $\int \hat{\rho}(\vec{v} | \vec{x},t) dv_1...dv_n = 1$. Therefore the problem becomes the following: Given a solution to the Schr\"{o}dinger equation represented by $\rho(\vec{x},t)$ and  $(b_1(\vec{x},t),...,b_n(\vec{x},t))$, construct an approximate solution $\hat{\rho}(\vec{x},\vec{v},t)$ (construct $\hat{\rho}(\vec{v}| \vec{x}, t)$)  to eq.$\ref{phaseprocessmulti}$ such that $\hat{\rho}(\vec{x},\vec{v},t)=\hat{\rho}(\vec{v}|\vec{x},t)\rho(\vec{x},t)$, $b_i(\vec{x},t) = \int v_i \hat{\rho}(\vec{v}|\vec{x}, t) dv_1...dv_n$ and $\int \hat{\rho}(\vec{v} | \vec{x},t) dv_1...dv_n = 1$. To construct a solution, choose the initial conditional density $\hat{\rho}(\vec{v}|\vec{x}, t=0)$ such that 

\begin{equation} \label{velocityav1}
\int v_i \hat{\rho}(\vec{v}| \vec{x}, t=0) dv_1...dv_n = b_i(\vec{x}, t=0)
\end{equation}
for each $i$. Assuming Conjecture 3 is true, $\vec{x}$ projections of the solutions to eq.$\ref{phaseprocessmulti}$ are close to a quantum process $\vec{x}_q(t)$ satisfying both of Nelson's laws. Therefore eq.$\ref{phaseprocessmulti}$ approximately propagates the initial $(\rho(\vec{x}, t=0), b_1(\vec{x}, t=0),...,b_n(\vec{x},t=0))$ to $(\rho(\vec{x}, t), b_1(\vec{x}, t),...,b_n(\vec{x},t))$ according to the Schr\"{o}dinger equation. Since $(\rho(\vec{x}, t=0), b_1(\vec{x}, t=0),...,b_n(\vec{x},t=0))$ completely specifies the initial conditions for the Schr\"{o}dinger equation (up to an additive constant in the initial phase $S(\vec{x},t=0)$), the solutions are unique. Therefore $(\hat{\rho}(\vec{x},t), \hat{b}_1(\vec{x},t),...,\hat{b}_n(\vec{x},t))=(\rho(\vec{x},t), b_1(\vec{x},t),...,b_1(\vec{x},t))$. Hence given a solution to the Schr\"{o}dinger equation specified by $\rho(\vec{x},t)$ and $(b_1(\vec{x},t),...,b_n(\vec{x},t))$, we see that any initial density $\hat{\rho}(\vec{x}, \vec{v}, t=0)$ satisfying eq.$\ref{velocityav1}$ and $\int \hat{\rho}(\vec{x}, \vec{v}, t=0)dv_1...dv_n=\rho(\vec{x},t=0)$ yields an approximate solution $\hat{\rho}(\vec{x}, \vec{v}, t)$ to eq.\ref{phaseprocessmulti} such that $\int v_i \hat{\rho}(\vec{v}| \vec{x}, t) dv_1...dv_n = b_i(\vec{x}, t)$  for each $i$ and $\int \hat{\rho}(\vec{x}, \vec{v}, t)dv_1...dv_n=\rho(\vec{x},t)$. Note that there are infinitely many choices of the initial conditional density $\hat{\rho}(\vec{v}| \vec{x}, t=0)$ compatible with eq.\ref{velocityav1}.

\section{Locality and local causality} \label{localitysection}
\subsection{Nonlocality of Nelson's theory vs locality of phase space model}

The multi-particle Nelson's theory is manifestly non-local as is well known \cite{Nelson2}. It is easy to see that the drift velocity $b_i$ of each particle depends on the positions of all other particles in general. Even if the particles are no longer interacting, a potential acting on one particle affects the drift velocities of other particles. In this sense, Nelson's formulation is non-local. As argued by Nelson in \cite{Nelson2} this could be due to the Markovian nature of noise and in a non-Markovian theory non-locality may not be required as non-locality could be traded off with temporal correlations. The phase space model presented here is non-Markovian due to the colored nature of the noises introduced. Is it also local? Assume that Conjectures 3 and 4 hold. The following simple argument is in favor of locality. Let the particles interact for some finite time until $t=0$ and we turn off the interactions (keeping single particle potentials) and separate the particles. Close to the white noise limit each particle obeys the equation

\begin{align} \label{phaseprocessmulti}
&dx_i(t) = v_i (t) dt + \epsilon A_i (t) dt \\ \nonumber	
&dv_i(t) = a_i(x_i(t))dt \\ \nonumber
\end{align}
As $A_i$'s will be uncorrelated for large $\beta_i$'s, every particle is subject to independent forces and independent potentials. The model is local in this sense. However the initial correlations can be preserved and this would give rise to quantum correlations. When the velocities are integrated out, we would obtain Nelson's theory which is non-local. The non-locality arises from integrating out degrees of freedom. We know from Bell type inequalities that certain theories cannot simulate quantum theory. Assuming that the measurement settings are independent from the variables of the theory, since the phase space is local, if it is simulate quantum theory, it cannot be locally causal \cite{bell_aspect_2004}.  In the next sections, I elaborate on this: local stochastic theories need not to be locally causal, therefore Bell-type inequalities provide no obstacle against the existence of a local stochastic underlying theory for quantum mechanics which is itself not locally causal.

\subsection{Multi-time measurements in stochastic mechanics}

In order to make contact with Bell-type inequalities, we need to review how to make sense of multi-time measurements in Nelson's theory. It is well known that for position measurements at a single time, the predictions of quantum mechanics and that of stochastic mechanics agree \cite{Nelson66}. It is quite easy to see this is the case. The quantum mechanical expectation value of functions $f$ and $g$ of the positions $x_1$ and $x_2$ at time $t$ is given by 

\begin{equation}
\int \rho(x_1, x_2, t) f(x_1) g(x_2)dx_1 dx_2,
\end{equation}
which is the same as that given by stochastic mechanics. Since the Schr\"{o}dinger equation and stochastic equations agree on the evolution of $\rho(x_1, x_2, t)$, the quantum expectation value is equal to the classical expectation value, therefore the theories are completely equivalent when position measurements are considered at a single time. However, it is also well known that the predictions of stochastic mechanics differ from that of quantum mechanics for measurements of commuting position variables at different times. The quantum mechanical expectation value of functions $f$ and $g$ of positions $x_1$ at time $t_1$ and $x_2$ at time $t_2$ is given by 

\begin{equation} \label{qmexpectation}
\int \rho(x_1, x_2, t_1) \rho(\tilde{x}_1, \tilde{x}_2, t_2 | x_1, x_2, t_1 ) f(x_1) g(\tilde{x_2})dx_1 dx_2 d\tilde{x}_1 d\tilde{x}_2
\end{equation}
where $\rho(\tilde{x}_1, \tilde{x}_2, t_2 | x_1, x_2, t_2 )$ is the quantum mechanical probability obtained from solving the Schr\"{o}dinger equation, for the particles reaching to $(\tilde{x_1}, \tilde{x}_2) $, at time $t_2$ given that their positions were $(x_1, x_2)$ at time $t_1$. In general, this multi-time expectation value contradicts the expectation value $\langle x_1(t_1) x_2(t_2) \rangle$ obtained in stochastic mechanics. All these generalize to multiple particles and measurements at more than $2$ times, or multi-time measurements.

For multi-time measurements, in order to make the quantum mechanical expectation values compatible with the expectation values calculated from stochastic mechanics, Blanchard et al \cite{blanchard} proposed that when the stochastic mechanical expectation values are computed, the probability density must be updated upon measurement. This process is the stochastic mechanical counterpart of the wave-function collapse. To be concrete, consider the two particle example. Assume that $t_2 > t_1$. When $x_1$ is measured at time $t_1$, the following reduction takes place in the probability density: $\rho(x_1, x_2, t_1)$ is replaced by $\rho_m(x_1, x_2, t_1 |\bar{x}_1)=\rho(x_2 | x_1, t_1) \delta(x_1 - \bar{x}_1)$, where $\bar{x}_1$ is the measured value of $x_1$. The stochastic equations are propagated with the new initial density $\rho_m(x_1, x_2, t_1|\bar{x}_1 )$ to $\rho_m(x_1, x_2, t_2 | \bar{x}_1)$. When the expectation values are computed, one first compute averages fixing $\bar{x}_1$, then averages over the density $\rho(\bar{x}_1, t_1)=\rho(x_1 = \bar{x}_1, t_1)$. This gives the quantum mechanical expectation value (eq. \ref{qmexpectation}) \cite{blanchard}. However, if one does not take into account the probability density reduction and propagates the initial density $\rho(x_1, x_2, t_2)$, one would get correlations in stochastic mechanics which would contradict the quantum expectation values. It must be stressed that this property is very peculiar to stochastic mechanics. In a usual diffusion process which is governed by a partial differential equation linear in probability, collapsing the initial probability, propagating the solution to a later time and averaging over the initial probability gives the same results for the statistics when the initial density is directly propagated. However, in stochastic mechanics, the drift velocities $b_i$ depend on the probability density, hence the diffusions are non-linear. When the probability is updated upon measurement and propagated, the drift velocities would be different from the the ones obtained by propagating the initial density directly. It is this non-linearity of stochastic mechanics seems to enable the probability collapse to make the quantum and stochastic mechanical expectation values equal for multi-time position measurements. For details, see \cite{blanchard}. 

Is the above probability collapse scheme non-local? Assume Conjecture 3 and 4 hold. Nelson's stochastic mechanics is manifestly non-local: anything done on the first particle will be felt by the second particle instantly because of the drift $b_2$ of the second particle depends on the position $x_1$ of the first particle in a general entangled state. The non-locality of Nelson's mechanics emerges from integrating out the velocity variables of the phase model, which is local. If one takes the point of view that the probabilities only represent the knowledge that one has about the underlying particle trajectories, there is nothing non-local in terms of how the particles evolve. It is the probability distribution that collapses, which is not part of ontology. The real quantities, which are the phase space stochastic trajectories of the particles, are not affected by each other, as is clear from that they are dynamically uncoupled and the random forces driving them are independent.

How to make contact with Bell type theorems? Consider the two particle case. Suppose that Albert has the two particles with phase coordinates $(x_1, v_1)$ and $(x_2, v_2)$ and he prepares them in the state $\rho_0(x_1, v_1, x_2, v_2)$, which corresponds to an entangled quantum states, by some local procedure. Then he sends the first particle to John and the second particle to Edward. Let John and Edward synchronize their clocks. John makes a measurement of position $x_1$ with outcome $\bar{x}_1$ at time $t_1$ and Edward makes a measurement of position $x_2$ with outcome $\bar{x}_2$ at a later time $t_2$. After this procedure is repeated many times, John and Edward report back their measurement outcomes to Albert for him to calculate empirical expectation values of functions of $\bar{x}_1$ and $\bar{x}_2$. Since $\rho(x_1, x_2,t)$ evolves according to the Schr\"{o}dinger equation, and since the quantum Heisenberg position operators $X_1(t_1)$ and $X_2(t_2)$ commute, the quantum mechanical expectation value $\langle f(X_1(t_1)) g(X_2(t_2))\rangle$ is given by eq.$\ref{qmexpectation}$. This must coincide with the empirical expectation values that Albert has computed since the particles obey the Schr\"{o}dinger equation. How would Albert compute expectation values in the stochastic model? He has two major choices. Either for each experiment, he assumes that John measured $x_1$ at $t_1$ and it yielded $\bar{x}_1$  and he updates the probability density to make a refined estimate of Edward's measurement of $x_2$ at $t_2$ conditioned on $\bar{x}_1$, then he averages over the probability density $\rho(\bar{x}_1)$, or he calculate the expectations directly without collapsing the probability distribution.  We discussed earlier in this section that these would in general give different results. If Albert chooses to update the probability density of two particles upon measurement, then he would obtain the quantum mechanical expectation values. Therefore, we assume that Albert calculate the expectation values in this way to make sure that multi-time measurements in quantum theory and stochastic mechanics coincide without violating locality.

\subsection{Local causality}

With the above measurement prescription, we obtain a correspondence between stochastic mechanics and quantum theory with multi-time measurements. Assuming Conjectures 3 and 4 are true and the measurement settings are independent from the particle variables (non-conspiring), this provides a local stochastic theory for quantum theory. Bell-type theorems \cite{bell_aspect_2004} imply that local, locally causal and non-conspiring deterministic or stochastic hidden variable theories satisfy inequalities that are violated by quantum mechanics. As quantum mechanics is not locally causal \cite{bell_aspect_2004}, neither the phase space model can be if the conjectures 3 and 4 are true.

Although local deterministic theories are locally causal, these two versions of locality are separate for stochastic theories. There is no reason that a local stochastic theory be locally causal automatically. See chapter 23 of \cite{Nelson2}. To elaborate on this, lets remind ourselves what locality and local causality means in the sense of Bell \cite{bell_aspect_2004}. Locality means that influences cannot propagate with a speed greater than that of light. This is the usual relativistic notion of locality. Local causality is subtler. Let $R1$ and $R2$ be spacelike separated regions. Let $A$($B$) be stochastic variables (beables) in $R1$($R2$). Denote the beables of the past of $R1$ ($R2$) minus the intersection of pasts of $R1$ and $R2$ as $N$($M$). Denote the beables in the intersections of the pasts of $R1$ and $R2$ as $\Lambda$. Local causality states that $p(A | \Lambda, N, B ) = p(A | \Lambda, N)$: the beables at $R2$ does not affect the outcomes at space-like seperated $R1$ given the past beables $\Lambda$ and $N$ of $R1$. This is of course true in deterministic local theory as the past of $R1$ determines what happens at $R1$ completely. In a stochastic theory where the dynamics is deterministic and initial conditions are unknown, local causality still holds. However, in general local causality is independent from locality for stochastic theories \cite{Nelson2}. A sufficiently chaotic theory can cause loss of information from the past of $R1$ to $R1$, still retaining useful information in $R2$ as correlations. This chaotic dynamics can emerge from deterministic dynamics of many degrees of freedom or it could be due to external noise. For a general stochastic theory, information in $R2$ could be non-redundant given the past of $R1$ without any causal connection between $R1$ and $R2$. Hence if conjectures 3 and 4 are to be true, the local phase space model cannot be locally causal.

\section{Fields} \label{sectionfields}
\subsection{Potential physical sources of noise}

How physical is to assume that the noise appears as velocity rather than a force in the phase space process? A possible answer to this comes from consideration of canonical momentum  rather than kinematic momentum. A similar problem appears when one quantize a charged non-relativistic particle in an electromagnetic field. We know that canonical commutation relations should be imposed on canonical rather than kinematic variables. Consider the Lagrangian of a unit charged particle with unit mass with position and velocity $(x,v)$ in an electromagnetic field $(\phi, A)$:

\begin{equation}
L = \frac{1}{2} v^2 - \phi(x,t ) + v \cdot A  
\end{equation}
The canonical momentum is

\begin{equation}
p = \frac{\partial L}{\partial v_i} = v_i + A_i
\end{equation}
and the equation of motion is 

\begin{equation}
\dot{p_i} = \frac{\partial L}{\partial x_i} = -\frac{\partial \phi}{\partial x_i} = a_i
\end{equation}
With the canonical variables $x, p$ we have 

\begin{align}
& dx_i =  (p_i - A_i) dt \\ \nonumber
& dp_i = a_i dt
\end{align}
Restricting to one dimension we have

\begin{align}
& dx =  p dt-A dt\\ \nonumber
& dp = a(x) dt
\end{align}
Now we see that if the electromagnetic potential $A$  is the colored noise $A_\beta$, we have induced noise as velocity. On the other hand if we have used kinematic momentum $v$, we would not have the noise term as velocity but acceleration. Having the noise term as a velocity seems to be important to make direct contact with Nelson's first law. 

A similar argument can be given for a non-relativistic particle in a weak gravitational field where again the canonical and kinematic momenta differ and the source of noise would be gravitational. In the next sections, I provide a quantization of the free scalar field in a stochastic gravitational field assuming conjectures 3 and 4. 

\subsection{Stochastic quantization of the free scalar field}

The phase space model for multiple particles enables a generalization to fields. The simplest case is the free scalar field though interacting fields and fields with spin could be treated in similar ways with some technical complications. First, I review stochastic quantization of the free scalar field following \cite{Guerra}. Consider the Klein-Gordon field $\phi$ satisfying 

\begin{equation} \label{KleinGordon}
\partial_t^2 \phi - \Delta \phi + m^2 \phi =0
\end{equation}
where $\Delta$ is the three dimensional Laplace operator. Put the system in a box $B$ with appropriate boundary conditions with a real complete orthonormal basis of eigenfunctions $u_i$ of $\Delta$ satisfying

\begin{equation}
\int_B u_i(x) u_j(x) d^3 x = \delta_{ij}
\end{equation}
where $\sum_i u_i(x) u_i (x') = \delta_B(x-x')$ (delta function in $B$: $\delta_B(x-x') \rightarrow \delta(x-x')$ in the infinite volume limit) and $\Delta u_i(x) = -k_i^2 u_i(x)$. Expand $\phi$ in the form

\begin{equation} \label{decomp}
\phi(x,t) =\sum_i q_i(t) u_i(x)
\end{equation}
or
\begin{equation} \label{decompinv}
q_i(t) = \int_B \phi(x,t) u_i(x) d^3x.
\end{equation}
Then each $q_i(t)$ is a harmonic oscillator with frequency $\omega_i=\sqrt{m^2+k_i^2}$:

\begin{equation}
\frac{d^2 q_i}{dt^2} + \omega_i^2 q_i=0.
\end{equation}
Assume the Hamiltonian of the field is $\int \frac{1}{2}( (\partial_t \phi)^2 + \partial_i \phi \partial^i \phi + m^2 \phi^2)d^3x$. To stochastically quantize the field, we promote each of the harmonic oscillators to quantum oscillators unit mass (not to be confused with the field mass $m$):

\begin{equation} \label{Nelsonfirstfield}
dq_i(t) = b_i(q_1(t), ..., q_n(t), t) dt + \epsilon dW_i(t)
\end{equation}
and
\begin{equation} \label{NNfield}
\frac{1}{2} (D_+ D_- + D_- D_+) q_i(t) = -\omega_i^2 q_i(t).
\end{equation}
where $dW_i(t)$'s are independent Wiener processes and $\epsilon = \sqrt{\hbar}$. Does the noise $\{dW_i(t)\}_i$ act locally on $\phi$? To see this, calculate the stochastic derivative of $\phi$ using eq.\ref{decomp}:

\begin{equation}
d\phi(x,t) = \sum_i b_i(q_1(t), ..., q_n(t), t) u_i(x) dt + \epsilon \sum_i u_i (x)dW_i(t)
\end{equation}
for each $x \in B$. The noise term is again Gaussian since it is a linear sum of Gaussian processes with mean and covariance

\begin{equation}
\epsilon \langle \sum_i u_i(x)dW_i(t) \rangle =0
\end{equation}

\begin{equation}
\epsilon^2 \langle \sum_i u_i(x) dW_i(t) \sum_j u_j(x') dW_j(t) \rangle = \epsilon^2\sum_i u_i(x) u_i(x') dt = \epsilon^2 \delta(x-x')dt
\end{equation}
in the infinite volume limit where we used $\langle dW_i(t)\rangle =0$ and $\langle dW_i(t) dW_j(t) \rangle = \delta_{ij}dt$. Hence, we can write

\begin{equation}
d\phi(x,t) = \sum_i b_i(q_1(t), ..., q_n(t), t) u_i(x) dt + \epsilon dW(x,t).
\end{equation}
with $\langle dW(x,t) \rangle =0$ and $\langle dW(x,t) dW(x', t)\rangle=\delta(x-x')dt$. In this sense, the noise acts locally on $\phi(x,t)$. Note that each $b_i(q_1(t), ..., q_n(t), t)$ can be written in terms of $\phi(x,t)$ using eq.\ref{decompinv}. Nelson's second law has a simple expression in terms of for the field:

\begin{equation} \label{Nelsonsecondfield}
\frac{1}{2} (D_+ D_- + D_- D_+) \phi(x,t) = ( \Delta  - m^2 ) \phi(x,t)
\end{equation}
as can be seen using eq.\ref{decomp} and eq.\ref{NNfield}. This can be also obtained by replacing the acceleration term $\partial_t^2 \phi$ by its stochastic counterpart $1/2(D_+ D_- + D_- D_+) \phi$ in eq.\ref{KleinGordon}. Does this stochastic quantization depend on the time slicing? Yes, it is obviously dependent on coordinates. For a different time slicing, one gets a different stochastic process. All these stochastic processes together should preserve the relativistic covariance of the resulting quantum theory \cite{Guerra}.

Now, I give a phase space model for the stochastic field. Instead of eq.\ref{Nelsonfirstfield}, we have

\begin{align}
&dq_i = v_i dt + \epsilon A_{i}dt \\ \nonumber
&dv_i = -\omega_i^2 q_i dt \\ \nonumber
&dA_i = -\beta_i A_i dt + \beta dW_i
\end{align}
as we had for the multi-particle case. Assuming Conjectures 3 and 4 are true, for large $\beta$, this provides a phase space model for the stochastic field. Is this model local? To see this, write the above equations in terms of $\phi(x,t)$ using eq.\ref{decomp} and defining $V(x,t) = \sum_i v_i(t) u_i(x)$:

\begin{align} \label{phasemodelfield}
&d\phi(x,t) = V(x,t) dt + \epsilon \sum_i u_i(x) A_i(t) dt\\ \nonumber
&dV(x,t) = ( \Delta  - m^2 ) \phi(x,t)dt \\ \nonumber
&dA_i = -\beta_i A_i dt + \beta dW_i
\end{align}
for each $x \in B$. In the large $\beta_i$ limit, assuming $A_i$'s should become independent Wiener processes. Thus, as above, $\sum_i u_i(x) A_i(t)$ converges to a Wiener process with zero mean and covariance $\delta(x-x')dt$ in the infinite volume limit. The dynamical equation $dV(x,t) = ( \Delta  - m^2 ) \phi(x,t)dt$ is already local. Locality of $d\phi(x,t) = V(x,t) dt + \epsilon \sum_i u_i(x) A_i(t)$ is obtained in the infinite volume and large $\beta_i$ limit.

\subsection{Gravitational origins of noise}
Can the source of noise be gravitational? I outline a model below. Consider the scalar field $\phi$ in a weak stochastic background gravitational field $g_{\mu\nu}$ with the action

\begin{equation}
S = \frac{1}{2} \int d^4 x \sqrt{-g} (g_{\mu \nu} \partial^\mu \phi \partial^\nu \phi -m^2 \phi^2 - 2 \xi R \phi).
\end{equation}
where $R$ is the Ricci scalar, $g=\text{det}g$ and $\eta_{\mu \nu}=\text{diag}(1,-1,-1,-1)$. Putting $g_{\mu\nu}=\eta_{\mu\nu}+h_{\mu\nu}$, expanding $S$ to first order in $h_{\mu \nu}$ and integrating the non-minimal coupling term $\xi R \phi$ by parts we have

\begin{equation}
S = \frac{1}{2} \int d^4 x [\partial^\mu \phi \partial_\mu \phi - m^2 \phi^2 - h^{\mu\nu} (\partial_\mu \phi \partial_\nu \phi -\frac{1}{2}\eta_{\mu \nu} \partial^\sigma \phi \partial_\sigma \phi)  + 2 \xi (\partial_\nu h^{\mu\nu}-\partial^\mu h^\sigma_\sigma)\partial_\mu \phi]
\end{equation}
The canonical momentum of the field is 

\begin{equation}
\Pi = \frac{\delta S}{\delta (\partial^0 \phi)} = \partial_0 \phi - h_{0 \nu} \partial^\nu \phi + \frac{1}{2} h^\sigma_\sigma \partial_0 \phi + \xi (\partial^\nu h_{0 \nu}-\partial_0 h).
\end{equation}
In the quantum regime assume that the minimal coupling part can be neglected as compared to the non-minimal coupling part. Of course for the experimentally tested classical regime, the minimal coupling part should dominate the non-minimal part. We have
in the quantum regime
\begin{equation}
\partial_0 \phi  = \Pi - \xi (\partial^\nu h_{0 \nu}-\partial_0 h).
\end{equation}
The acceleration equation can be obtained from the Euler-Lagrange equation yielding

\begin{equation}
\partial_0 \Pi = \Delta \phi - m^2 \phi + F^{\mu \nu} (\phi, \partial_\sigma \phi) h_{\mu \nu}
\end{equation}
where $F^{\mu \nu}$ is some linear differential operator depending on $\phi$ and $\partial_\sigma \phi$.  In the Newtonian gauge $h_{\mu \nu} = \text{diag}(\theta, \theta, \theta, \theta)$ assuming some isotropy condition on the stress-energy tensor generating it. Hence, we have the system of equations

\begin{align} \label{phasemodelgrav}
& \partial_0 \phi  = \Pi + \xi \partial_0 \theta \\ \nonumber
& \partial_0 \Pi = \Delta \phi - m^2 \phi + G(\phi, \partial_\sigma \phi) \theta.
\end{align}
In order to stochastically quantize the field, we need to make the identification (assuming conjectures 3 and 4): $\xi \partial_0 \theta = \epsilon A_\beta$ where $A_\beta$ is a space-time colored noise, which is characterized by some set of parameters $\beta$, converging to space-time white noise (time derivative of the zero mean Wiener process with covariance $\delta(x-x')dt$ as above in the infinite volume limit) as $\beta$ becomes large. Assuming that the statistics of $\theta$ becomes independent of $\phi$ when $\beta$ is large and $\langle \theta \rangle =0$, eq.\ref{phasemodelgrav} yields Nelson's second law (eq.\ref{Nelsonsecondfield}). This finalizes the sketch of a phase space quantization of the scalar field in a stochastic background gravitational field. Note that, this cannot be achieved with minimal coupling in the framework considered here since the noise terms induced from minimal coupling would depend on the field itself. 

Can this background gravitational field be sourced by a matter distribution? In the limit of large $\beta$, in the infinite volume limit, $\theta$ is characterized by a space-time Gaussian field with zero mean and covariance:

\begin{equation}
\langle d\theta(x, t) d\theta(x', t) \rangle = \frac{\epsilon^2}{\xi^2} \delta(x-x') dt
\end{equation}
In a Newtonian universe, assume small fluctuations around a constant mean (dark) matter density. A zero mean homogeneous isotropic Gaussian matter distribution can be described by the two-point function of the fluctuating part of the matter density $\delta\rho(x)$:

\begin{equation}
c(x-x',t)= \langle \delta\rho (x, t) \delta\rho (x', t)\rangle
\end{equation}
whose Fourier transform $P(k)$ is a power law:

\begin{equation}
P(k,t) = \int c (x,t) e^{-i k \cdot x} d^3x = A(t) |k|^{n}.
\end{equation}
The spectrum of the gravitational potential $\theta$ can be calculated from the Poisson equation:

\begin{equation}
\Delta \phi  = 4 \pi G \delta\rho
\end{equation}
where $G$ is the Newton's constant. Taking the Fourier transform of both sides we can show that the spectrum $P_{\theta}(k)$ of $\theta$ is given by

\begin{equation}
P_\theta (k, t) = \frac{(4 \pi G )^2}{|k|^4} P(k,t).
\end{equation}
The Fourier transform of the covariance of $\theta$ is:
\begin{equation}
P_\theta(k,t) =  \frac{\epsilon^2}{\xi^2} t
\end{equation}
from which we obtain
\begin{equation}
P(k,t) = \frac{|k|^4}{(4 \pi G )^2} \frac{\epsilon^2}{\xi^2} t.
\end{equation}
Analogous expressions can be written for the case of expanding universe. Of course, this is a tall order: fluctuations should remain up until where quantum field theory is tested in particle accelerators. Also, Planck's constant is emergent and can depend on time in principle. Note that the matter distribution should temporally fluctuate as white noise. This is consistent assuming that the fluctuations arise from the gravitational interactions of all particles (fields) in the universe as Brownian motion of single particles emerge from the microscopic dynamics in a gas or plasma \cite{calogero}.

\bibliographystyle{unsrt}
\bibliography{derivation_of_nelson4_references}

\end{document}